\documentclass[reprint, aps, prb]{revtex4-1}
\usepackage{amsfonts}
\usepackage{amssymb}
\usepackage{amsmath}
\usepackage{graphics}
\usepackage{graphicx}
\usepackage{subfigure}
\usepackage{dcolumn}
\usepackage{bm}
\usepackage{hyperref}
\usepackage{color}
\usepackage{epstopdf}

\DeclareMathAlphabet\mathbfcal{OMS}{cmsy}{b}{n}

\begin{document}

\title{Graphene valley polarization as a function of carrier-envelope phase
in few-cycle laser pulses and its footprints in harmonic signals}
\author{H.K. Avetissian}
\author{V. A. Sedrakyan}
\author{Kh. V. Sedrakian}
\author{G. F. Mkrtchian}
\thanks{mkrtchian@ysu.am}

\affiliation{Centre of Strong Fields Physics, Yerevan State University,
Yerevan 0025, Armenia}

\begin{abstract}
We consider coherent dynamics of graphene charged carriers exposed to an
intense few-cycle linearly polarized laser pulse. The results, obtained by
solving the generalized semiconductor Bloch equations numerically in the
Hartree-Fock approximation, taking into account many-body Coulomb
interaction, demonstrate strong dependence of the valley polarization on the
carrier-envelope phase (CEP), which is interpolated by the simple sinusoidal
law. Then we consider harmonic generation in multi-cycle laser field by
graphene preliminary exposed to an intense few-cycle laser pulse. We show
that the second harmonic's intensity is a robust observable quantity that
provides a gauge of CEP for pulse durations up to two optical cycles,
corresponding to 40 $\mathrm{fs}$ at the wavelength of 6.2 $\mathrm{\mu m}$.
\end{abstract}

\maketitle

\section{Introduction}

In many semiconductor and semimetal materials, conduction band electrons
occupy states near several discrete energy minima which have been termed
valleys. In general, electrons in these discrete valley-states have
different properties, which results in a valley-dependent electromagnetic
(EM) response in these materials. Hence, in addition to electron spin, we
have an additional degree of freedom -- valley isospin or valley
polarization. In analogy with the spintronics for spin-based technology, it
has been proposed the valleytronics \cite{valley1,valley2,valley3} that has
attracted a considerable attention because of its possible application in
quantum information science. Experimentally the valley-polarization of
electrons was achieved in several materials. In AlAs, valley polarization
was induced by a symmetry-breaking strain \cite{AlAs}, in bismuth - by using
a rotating magnetic field \cite{bismuth}, in diamond -by electrostatic
control of valley currents \cite{diamond1,diamond2}, in MoS$_{2}$ -by means
of circularly polarized light due to valley-contrasting Berry curvature \cite%
{Cao,Mak,Zeng,Li}.

The valley polarization is sensitive to thermal lattice vibrations and for
valleytronics applications, it is necessary materials where the valley
polarization relaxation time is remarkably long. From this point of view
graphene \cite{Castro}, where the intervalley scattering is suppressed \cite%
{Morpurgo,Morozov,McCann}, is of interest. For graphene, there are two
inequivalent Dirac points in the Brillouin zone, related by time-reversal
and inversion symmetry \cite{Castro}. Hence, the valley-contrasting Berry
curvature is zero that makes it difficult to use graphene in valleytronics,
especially if one wants to use intense light pulses for ultrafast
manipulation of valley polarization. A number of proposals on how to break
time-reversal or inversion symmetry in graphene based materials for
generation of the valley polarization has been put forward. Valley
polarization in graphene sheet can be achieved with zigzag edges \cite%
{valley1}, at a line defect \cite{Gunlycke}, for strained graphene with
massive Dirac fermions \cite{strain1,strain2} and with broken inversion
symmetry \cite{Xiao,Yankowitz}, through a boundary between the monolayer and
bilayer graphene \cite{Nakanishi,Pratley}. Valley polarization in graphene
can also be achieved at the breaking time-reversal symmetry with AC
mechanical vibrations \cite{Jiang}. Regarding the light-wave manipulation of
valleys, it was shown that valley currents can be induced by asymmetric
monocycle EM pulses \cite{Moskalenko} or by specifically polarized light 
\cite{Golub}. Recently, for gapped graphene like systems, such as hexagonal
boron nitride and molybdenum disulfide several schemes for light-wave
manipulation of valleys have been proposed which do not rely on
valley-contrasting Berry curvature. One with two-color counter-rotating
circularly polarized laser pulses \cite{Ivanov1} and the other by exploiting
the carrier-envelope phase (CEP) of ultrashort linearly polarized pulse \cite%
{Ivanov2}, or circularly polarized pulse \cite{Motlagh}. Successful attempts
have been made to apply these schemes to intrinsic graphene \cite%
{Mrudul1,Mrudul2,Kelardeh}, where for the linear polarization of the driving
wave the main emphases have been made to subcycle laser pulses \cite%
{Kelardeh}.

The problem of coherent interaction of graphene with an intense few-cycle
linearly polarized laser pulse can be applied to solve two important issues
regarding the valleytronics and light-wave electronics \cite%
{goulielmakis2007attosecond}. For valleytronics, it can be an efficient
method to reach valley-polarization which depends on the carrier-envelope
phase, and vice versa, to measure CEP via the valley-polarization which is
important for short pulse manipulation. As is well known, the second
harmonic is sensitive to valley-polarization \cite{Golub2,Wehling,Ho}, which
opens a door for solving the mentioned issues. In the present paper, we
address these problems considering the coherent dynamics of graphene charged
carriers exposed to an intense few-cycle linearly polarized laser pulse. We
numerically solve the generalized semiconductor Bloch equations in the
Hartree-Fock approximation \cite{Knorr,Mer18,Mer2022} taking into account
the many-body Coulomb interaction and demonstrate the strong dependence of
the valley polarization on the CEP, which is interpolated by the simple
harmonic law. Then we consider the harmonic generation in a multi-cycle
laser pulse-field by graphene preliminary exposed to an intense few-cycle
laser pulse and consider the ways to gauge the CEP.

The paper is organized as follows. In Sec. II, the model and the basic
equations are formulated. In Sec. III, we present the main results. Finally,
conclusions are given in Sec. IV.

\section{The model and theoretical methods}

We begin by describing the model and theoretical approach. Two dimensional
hexagonal nanostructure is assumed to interact with a few cycle mid-infrared
laser pulse that excites coherent electron dynamics which is subsequently
probed by an intense near-infrared or visible light pulse generating high
harmonics. It is assumed that the polarization planes of both laser fields
coincide with the nanostructure plane ($XY$). For numerical convenience and
to avoid residual momentum transfer, the electric field $\mathbf{E}\left(
t\right) $ is calculated from the expression of the vector potential given
by 
\begin{equation*}
\mathbf{A}\left( t\right) =f_{0}\left( t\right) A_{0}\hat{\mathbf{e}}%
_{0}\sin \left( \omega _{0}t+\phi _{\mathrm{CEP}}\right)
\end{equation*}%
\begin{equation}
+f_{1}\left( t\right) A_{1}\hat{\mathbf{e}}_{1}\sin \left( \omega _{1}t+\phi
_{\mathrm{CEP}1}\right) ,  \label{Et}
\end{equation}%
where $A_{0}$ and $A_{1}$ are the amplitudes of the vector potentials which
are connected to the amplitudes of the applied electric fields $E_{0}=\omega
_{0}A_{0}$ and $E_{1}=\omega _{1}A_{1}$ of the laser pulses, $\omega _{0}$
and $\omega _{1}$ are the fundamental frequencies, $\hat{\mathbf{e}}_{0}$
and $\hat{\mathbf{e}}_{1}$ are the unit polarization vectors, $\phi _{%
\mathrm{CEP}}$ and $\phi _{\mathrm{CEP}1}$ are carrier-envelope phases. The
envelopes of the two waves are described by the sin-squared functions%
\begin{equation}
f_{0,1}(t+t_{0,1})=\left\{ 
\begin{array}{cc}
\sin ^{2}\left( \pi t/\tau _{0,1}\right) , & 0\leq t\leq \tau _{0,1}, \\ 
0, & t<0,t>\tau _{0,1},%
\end{array}%
\right.  \label{envelop1}
\end{equation}%
where $\tau _{0}$ and $\tau _{1}$ characterize the pulses' duration, and $%
t_{1}$ and $t_{2}$ define the starts of the pulses. For a probe wave we will
assume $\tau _{1}>>2\pi /\omega _{1}$. In this case the detailed behavior of
the electric field dependence on the carrier-envelope phase is not
important. Hence we take $\phi _{\mathrm{CEP}1}=0$. For a few-cycle pump
pulse the exact position of nodes and peaks of the electric field is of
great importance, since it determines the momentum distribution of the
excited electrons, which in turn defines harmonics spectrum during the
interaction with the probe pulse. For a pump wave-pulse the envelope $%
f_{0}(t)$ is chosen such that the maximal electric field strength is reached
at $t=0$ and $\phi _{\mathrm{CEP}}=0$.

Monolayer graphene interaction with the EM field is described using
generalized semiconductor Bloch equations in the Hartree-Fock approximation.
Technical details can be found in Refs. \cite{Knorr,Mer18,Mer2022}, while
for self-consistency the main equations are outlined below. Thus, the Bloch
equations within the Houston basis read 
\begin{equation*}
\partial _{t}\mathcal{N}(\mathbf{k}_{0},t)=-2\mathrm{Im}\left\{ \left[ 
\mathbf{E}\left( t\right) \mathbf{D}_{\mathrm{tr}}\left( \mathbf{k}_{0}+%
\mathbf{A}\right) \right. \right.
\end{equation*}%
\begin{equation}
\left. \left. +\Omega _{c}\left( \mathbf{k}_{0}+\mathbf{A},t;\mathcal{P},%
\mathcal{N}\right) \right] \mathcal{P}^{\ast }(\mathbf{k}_{0},t)\right\} ,
\label{1}
\end{equation}%
\begin{equation*}
\partial _{t}\mathcal{P}(\mathbf{k}_{0},t)=-i\left[ \mathcal{E}_{eh}\left( 
\mathbf{k}_{0}+\mathbf{A}\right) -i\Gamma \right] \mathcal{P}(\mathbf{k}%
_{0},t)
\end{equation*}%
\begin{equation*}
+i\left[ \mathbf{E}\left( t\right) \mathbf{D}_{\mathrm{tr}}\left( \mathbf{k}%
_{0}+\mathbf{A}\right) +\Omega _{c}\left( \mathbf{k}_{0}+\mathbf{A},t;%
\mathcal{P},\mathcal{N}\right) \right]
\end{equation*}%
\begin{equation}
\times \left[ 1-2\mathcal{N}(\mathbf{k}_{0},t)\right] ,  \label{2}
\end{equation}%
where $\mathcal{P}(\mathbf{k},t)=\mathcal{P}^{\prime }(\mathbf{k},t)+i%
\mathcal{P}^{\prime \prime }(\mathbf{k},t)$ is the interband polarization
and $\mathcal{N}\left( \mathbf{k},t\right) $ is the electron distribution
function for the conduction band. The electron-hole energy 
\begin{equation}
\mathcal{E}_{eh}\left( \mathbf{k}\right) =2\mathcal{E}\left( \mathbf{k}%
\right) -\Xi _{c}(\mathbf{k},t;\mathcal{P},\mathcal{N})  \label{21}
\end{equation}%
is defined via the band energy 
\begin{equation}
\mathcal{E}\left( \mathbf{k}\right) =\gamma _{0}\left\vert f\left( \mathbf{k}%
\right) \right\vert ,  \label{3}
\end{equation}%
and many-body Coulomb interaction energy%
\begin{equation*}
\Xi _{c}(\mathbf{k},t;\mathcal{P},\mathcal{N})=\frac{2}{\left( 2\pi \right)
^{2}}\int_{BZ}d\mathbf{k}^{\prime }V_{2D}\left( \mathbf{k-k}^{\prime }\right)
\end{equation*}%
\begin{equation}
\times \left\{ f_{c}\left( \mathbf{k,k}^{\prime }\right) \mathcal{N}\left( 
\mathbf{k}^{\prime }\right) +f_{s}\left( \mathbf{k,k}^{\prime }\right) 
\mathcal{P}^{\prime \prime }\left( \mathbf{k}^{\prime },t\right) \right\} .
\label{4}
\end{equation}%
In Eqs. (\ref{3}) $\gamma _{0}$ is the transfer energy of the
nearest-neighbor hopping and the structure function is 
\begin{equation}
f\left( \mathbf{k}\right) =e^{i\frac{ak_{y}}{\sqrt{3}}}+2e^{-i\frac{ak_{y}}{2%
\sqrt{3}}}\cos \left( \frac{ak_{x}}{2}\right) ,  \label{5}
\end{equation}%
where $a$ is the lattice spacing. In Eq. (\ref{4}) 
\begin{eqnarray*}
f_{c}\left( \mathbf{k,k}^{\prime }\right) &=&\cos \left[ \mathrm{arg}f\left( 
\mathbf{k}^{\prime }\right) -\mathrm{arg}f\left( \mathbf{k}\right) \right] ,
\\
f_{s}\left( \mathbf{k,k}^{\prime }\right) &=&\sin \left[ \mathrm{arg}f\left( 
\mathbf{k}^{\prime }\right) -\mathrm{arg}f\left( \mathbf{k}\right) \right] .
\end{eqnarray*}%
The electron-electron interaction potential is modelled by screened Coulomb
potential \cite{Knorr}: 
\begin{equation}
V_{2D}\left( \mathbf{q}\right) =\frac{2\pi }{\epsilon \epsilon _{\mathbf{q}%
}\left\vert \mathbf{q}\right\vert },  \label{6}
\end{equation}%
which accounts for the substrate-induced screening in the 2D nanostructure ($%
\epsilon $) and the screening stemming from valence electrons ($\epsilon _{%
\mathbf{q}}$). We take $\epsilon \simeq 4$ which is close to the value of a
graphene layer sandwiched by a SiO$_{2}$. The screening induced by
nanostructure valence electrons is calculated within the Lindhard
approximation of the dielectric function $\epsilon _{\mathbf{q}}$. \textrm{I}%
n general, one should consider the problem of dynamic screening. However,
the used here ansatz is justified since the large $q$\ behavior of graphene
dielectric screening does not depend on the electron density \cite%
{hwang2007dielectric} and we consider ultrashort time scales. In Eq. (\ref{2}%
) $\Gamma $ is the damping rate. In Eqs. (\ref{1}) and (\ref{2}) the
interband transitions are defined via the transition dipole moment%
\begin{equation*}
\mathbf{D}_{\mathrm{tr}}\left( \mathbf{k}\right) =-\frac{a}{2\left\vert
f\left( \mathbf{k}\right) \right\vert ^{2}}\sin \left( \frac{\sqrt{3}}{2}%
ak_{y}\right) \sin \left( \frac{ak_{x}}{2}\right) \widehat{\mathbf{x}}
\end{equation*}%
\begin{equation}
+\frac{a}{2\sqrt{3}\left\vert f\left( \mathbf{k}\right) \right\vert ^{2}}%
\left( \cos \left( ak_{x}\right) -\cos \left( \frac{\sqrt{3}}{2}%
ak_{y}\right) \cos \left( \frac{ak_{x}}{2}\right) \right) \widehat{\mathbf{y}%
},  \label{7}
\end{equation}%
and the light-matter coupling via the internal dipole field of all generated
electron-hole excitations: 
\begin{equation*}
\Omega _{c}\left( \mathbf{k},t;\mathcal{P},\mathcal{N}\right) =\frac{1}{%
\left( 2\pi \right) ^{2}}\int_{BZ}d\mathbf{k}^{\prime }V_{2D}\left( \mathbf{%
k-k}^{\prime }\right)
\end{equation*}%
\begin{equation}
\times \left\{ \mathcal{P}^{\prime }\left( \mathbf{k}^{\prime },t\right)
+if_{c}\left( \mathbf{k,k}^{\prime }\right) \mathcal{P}^{\prime \prime
}\left( \mathbf{k}^{\prime }\right) -if_{s}\left( \mathbf{k,k}^{\prime
}\right) \mathcal{N}\left( \mathbf{k}^{\prime },t\right) \right\} .
\label{8}
\end{equation}

For compactness of equations atomic units are used throughout the paper
unless otherwise indicated. The initial conditions $\mathcal{P}(\mathbf{k}%
,0)=0$ and $\mathcal{N}(\mathbf{k},0)=0$ are assumed, neglecting thermal
occupations. We will solve these equations numerically. It is more
convenient to make integration of these equations in the reduced BZ which
contains equivalent $\mathbf{k}$-points of the first BZ, cf. Fig. 1.

\begin{figure}[tbp]
\includegraphics[width=.35\textwidth]{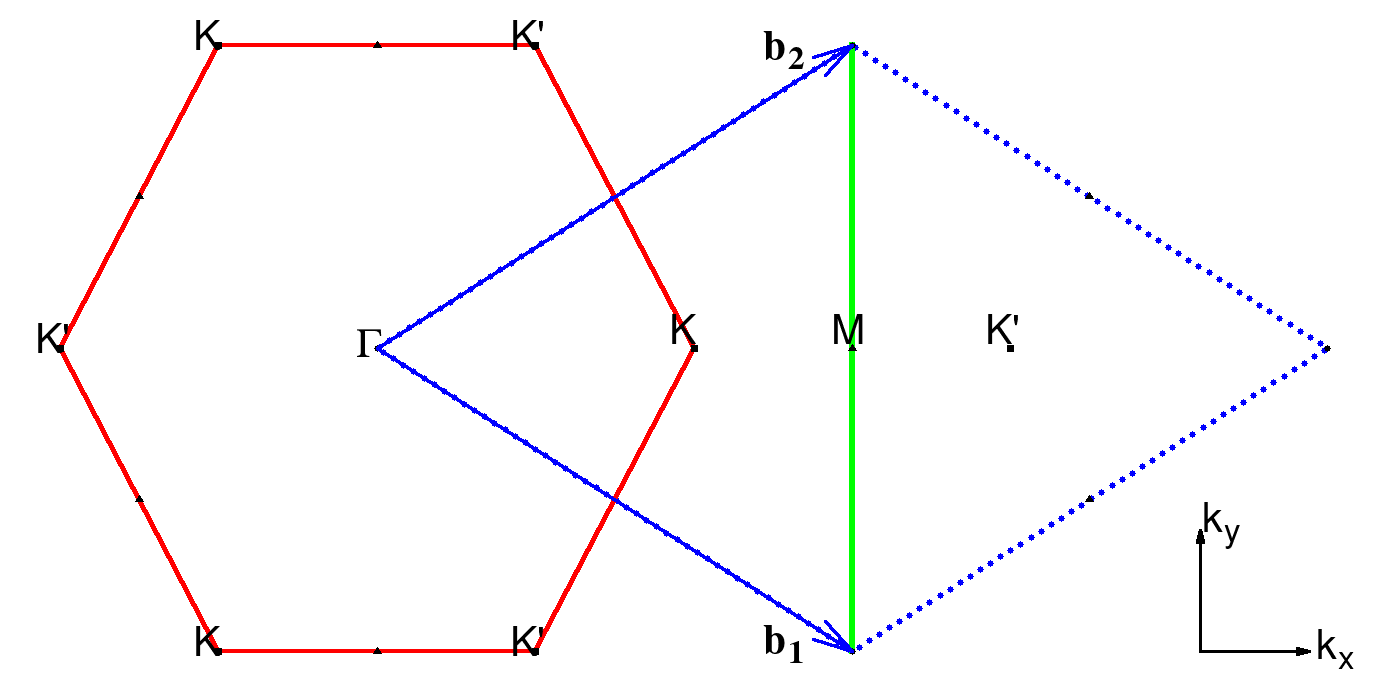}
\caption{The hexagonal first BZ of the reciprocal lattice (red solid line).
The rhombus (blue dashed lines) formed by the reciprocal lattice vectors is
a reduction of the second BZ and contains the same vectors of the first BZ.
The green (solid) line, that goes through M point, divides the rhombus into
the left and right triangles, which contain two valleys described by the $%
D_{3h}$ group.}
\end{figure}

\begin{figure*}[tbp]
\includegraphics[width=1.0\textwidth]{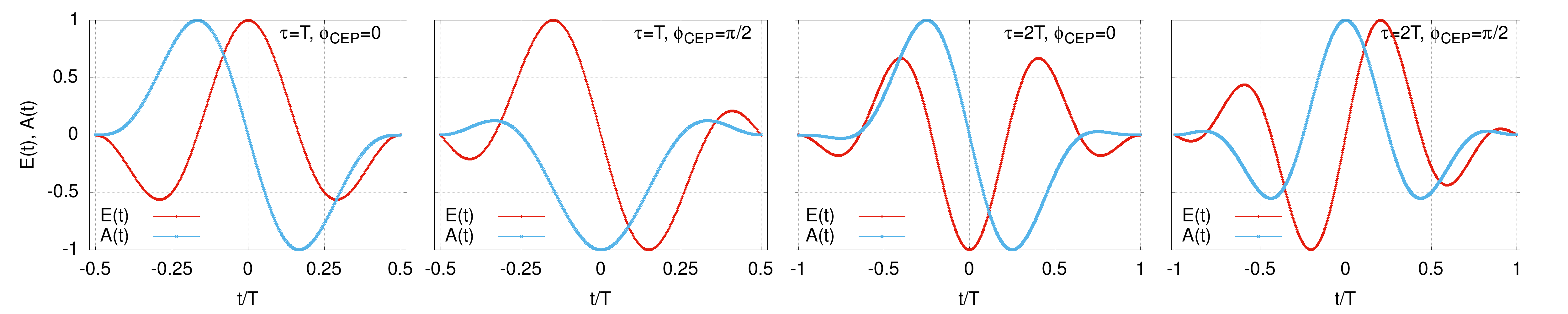}
\caption{The electric field and vector potental normalized to maximal values
versus time for one and two cycle pulses for a cosine pulse and for a sine
pulse.}
\end{figure*}

\begin{figure*}[tbp]
\includegraphics[width=1.0\textwidth]{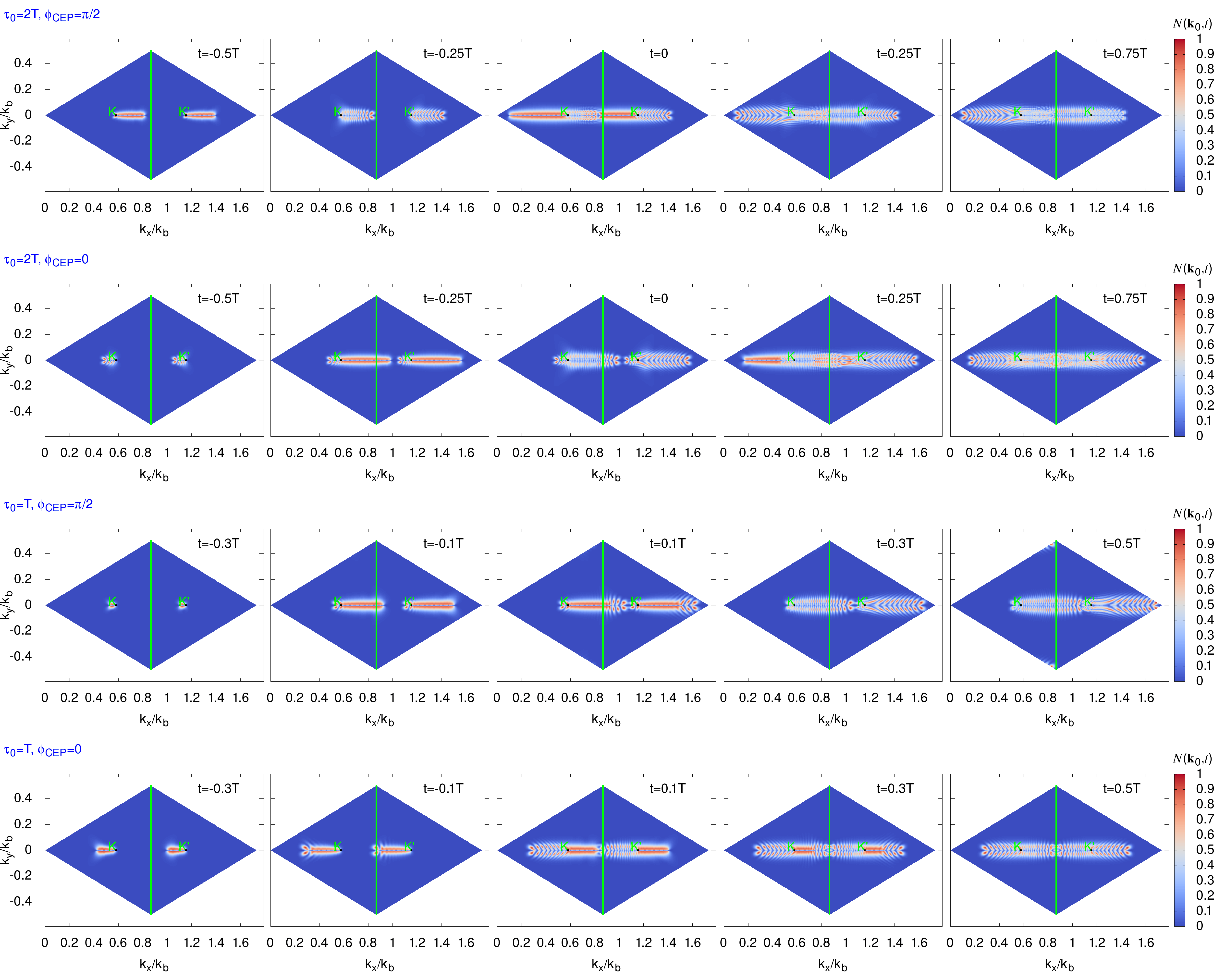}
\caption{ Particle distribution function $\mathcal{N}\left( \mathbf{k}%
,t\right) $ (in arbitrary units) at various time instances during the
interaction with the two (upper two rows) and single (3rd and 4th rows)
cycle laser pulse for graphene, as a function of scaled dimensionless
momentum components ($k_{x}/k_{b}$, $k_{y}/k_{b}$). The fundamental
frequency is $\protect\omega _{0}=0.2\ \mathrm{eV}/\hbar $ and the
wave-particle interaction parameter is taken to be $\protect\chi _{0}=3.5$
(intensity $1\ \mathrm{TW/cm}^{2}$).}
\end{figure*}

To quantify valley polarization we introduce valley-asymmetry
parameter: 
\begin{equation}
\eta =2\frac{N_{+}-N_{-}}{N_{+}+N_{-}},  \label{VP}
\end{equation}%
where $N_{-}$ and $N_{+}$ are electron populations around respective valleys
and are obtained by integrating over left ($BZ_{L}$) and right triangles ($%
BZ_{R}$) in Fig.1:%
\begin{equation}
N_{-}(t)=\frac{1}{\left( 2\pi \right) ^{2}}\int_{BZ_{L}}d\mathbf{k}\mathcal{N%
}\left( \mathbf{k},t\right) ,  \label{Nm}
\end{equation}%
\begin{equation}
N_{+}(t)=\frac{1}{\left( 2\pi \right) ^{2}}\int_{BZ_{R}}d\mathbf{k}\mathcal{N%
}\left( \mathbf{k},t\right) .  \label{Np}
\end{equation}%
It is clear that the denominator in Eq. (\ref{VP}) is the total electron
population in the conduction band $N(t)$:%
\begin{equation}
N(t)=N_{-}(t)+N_{+}(t)=\frac{1}{\left( 2\pi \right) ^{2}}\int_{BZ}d\mathbf{k}%
\mathcal{N}\left( \mathbf{k},t\right) .  \label{Nt}
\end{equation}

\section{Results}

Having outlined the set-up, we first consider coherent dynamics of graphene
charged carriers exposed to an intense few-cycle linearly polarized laser
pulse. Our results are obtained by solving the corresponding generalized
semiconductor Bloch equations (\ref{1}) and (\ref{2}) analytically in the
free carrier approximation and numerically in the Hartree-Fock approximation
taking into account many-body Coulomb interaction.

\subsection{\label{subsec:Quantum}Valley polarization dependence on CEP in
graphene}

As is evident from Eqs. (\ref{1}) and (\ref{2}) the excitation pattern of BZ
is defined by the vector potential of the pump pulse and to have overall
valley polarization one should break the symmetry of free standing graphene.
Although the crystal lattice of graphene is centrosymmetric, the valleys $K$
and $K^{\prime }$ are described by the wave vector group $D_{\mathrm{3h}}$
lacking the space inversion. These valleys are in the left and right
triangles of the rhombus in Fig. 1 and are connected with each other by the
space inversion $I$. Thus, the overall symmetry of free standing graphene $%
D_{6h}=D_{3h}\times I$ is centrosymmetric.

In Fig. 2 we plot the electric field $E\left( t\right) $ and vector
potential $A\left( t\right) $ of one and two cycle pulses for a cosine pulse 
$\phi _{\mathrm{CEP}}=0$ and for the sine pulse $\phi _{\mathrm{CEP}}=\pi /2$%
. As is seen from Fig. 2, under time reversal $t\rightarrow -t$ the electric
field of a cosine pulse remains invariant, while the field of a sine pulse
changes sign. For the vector potential $A\left( t\right) $ that defines the
excitation pattern of BZ the situation is opposite. The sine pulse has a
preferred orientation of the vector potential --the orientation of the
maximal field peak. The violation of symmetry has a striking manifestation
in the valley polarization. To have maximal valley polarization the
amplitude of the vector potential should be along $\Gamma -K$ direction and
close to magnitude of the wave vector separation of two Dirac points $%
\mathrm{K}\left( k_{b}/\sqrt{3},0\right) $ and $\mathrm{K}^{\prime }\left(
2k_{b}/\sqrt{3},0\right) $, where $k_{b}=4\pi /\sqrt{3}a$. \ The
wave-particle interaction will be characterized by the dimensionless
parameter $\chi _{0,1}=eE_{0,1}a/\hbar \omega _{0,1}$ which represents the
work of the wave electric field $E_{0,1}$ on a lattice spacing in the units
of photon energy $\hbar \omega _{0,1}$. The parameter is written here in
general units for clarity. The total intensity of the laser beam expressed
by $\chi _{0,1}$, can be estimated as:%
\begin{equation}
I_{\chi _{0,1}}=\chi _{0,1}^{2}\times \lbrack \hbar \omega _{0,1}/\mathrm{eV}%
]^{2}\times \left[ \mathrm{\mathring{A}}/a\right] ^{2}\times 1.33\times
10^{13}\ \mathrm{W\ cm}^{-2}.  \label{intensity}
\end{equation}%
The condition $A_{0}\sim \left\vert \mathrm{K-K}^{\prime }\right\vert \ $is
equivalent to $\chi _{0}\sim 4\pi /3$. For both waves we will restrict
interaction parameters to keep the intensities below the damage threshold 
\cite{Yoshikawa} $I_{\chi _{0,1}}<I_{\mathrm{dam}}\,\allowbreak \simeq 2\
TW/cm^{2}$ of graphene. The experiments \cite%
{lui2010ultrafast,breusing2011ultrafast} on graphene with the Ti:sapphire
laser showed that the carriers in graphene are well thermalized among
themselves during the period of light emission via the very rapid
electron-electron scattering. This rapid relaxation is compatible with
theoretical studies \cite{hwang2007inelastic,tse2008ballistic}, which
predict scattering times of tens of femtoseconds. Hence, during the
interaction with a few-cycle mid-infrared laser pulse that excites coherent
electron dynamics the relaxation time is taken to be equal to the wave
period $\Gamma ^{-1}=2\pi /\omega _{0}$. When excited graphene is
subsequently probed by an intense near-infrared or visible light pulse the
relaxation time is assumed to be $\Gamma ^{-1}=2\pi /\omega _{1}$. 

\begin{figure*}[tbp]
\includegraphics[width=1.0\textwidth]{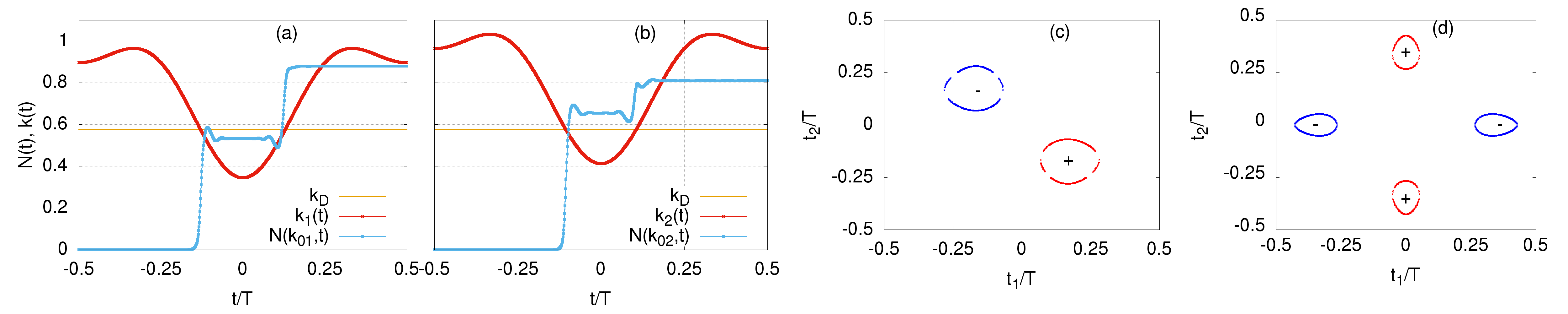}
\caption{Particle distribution function $\mathcal{N}\left( \mathbf{k}%
,t\right) $ at the interaction with a single cycle sine-pulse ($\protect\phi %
_{\mathrm{CEP}}=\protect\pi /2$) versus time for particular crystal momentum 
$\overline{\mathbf{k}}_{0}\mathbf{=}\left( k_{x}/k_{b},k_{y}/k_{b}\right) $:
(a) $\overline{\mathbf{k}}_{0}\mathbf{=}\left( 0.9,-0.033\right) \ $and (b) $%
\overline{\mathbf{k}}\mathbf{=}\left( 0.96,0.025\right) $. In these figures,
the x-component of kinetic momentum $\mathbf{k}\left( t\right) =\mathbf{k}%
_{0}+\mathbf{A}$ along with Dirac point momentum ($k_{D}/k_{b}=1/\protect%
\sqrt{3}$) are also plotted. All plotted quantities are dimensionless. In
(c) and (d) the roots of Eqs. (\protect\ref{minus}) and (\protect\ref{plus})
are plotted for cosine and sine pulses, respectively. The fundamental
frequency is $\protect\omega _{0}=0.2\ \mathrm{eV}/\hbar $ and the
wave-particle interaction parameter is taken to be $\protect\chi _{0}=4$
(intensity $1.4\ \mathrm{TW/cm}^{2}$). }
\end{figure*}

A typical coherent dynamics of graphene-charged carriers exposed to an
intense few-cycle linearly polarized laser pulse is shown in Fig. 3. The
laser field is polarized along $\Gamma -K$ direction (x-axis). In this
figure particle distribution function $\mathcal{N}\left( \mathbf{k},t\right) 
$ at various time instances during the interaction with the two and
single-cycle laser pulse is shown in reduced BZ. As indicated in this figure
one can trace the correlation between excitation patterns and time
dependence of the vector potential (Fig. 2). In particular, for sine pulses,
cf. first and third rows, thanks to the preferred direction of the vector
potential (cf. Fig. 2), one of the valleys is eventually populated. For
cosine pulses, cf. second and fourth rows, despite the valley polarization
during the first half of the laser pulse, the balance is recovered in the
second half of the laser pulse. Note that this picture of valley
polarization works only when the polarization of the laser is along $\Gamma
-K$ direction and it is of a threshold nature: the electrons created near
the one valley should pass to the opposite valley. To have a quantitatively
more satisfactory explanation we need to solve analytically Eqs. (\ref{1})
and (\ref{2}). To this end, we omit the Coulomb and relaxation terms in
Bloch equations Eqs. (\ref{1}) and (\ref{2}), which is justified for a
few-cycle laser pulse.

Thus, the formal solution of Eq. (\ref{2}) for the interband polarization
can be written as: 
\begin{equation*}
\mathcal{P}(\mathbf{k}_{0},t)=i\int_{-\tau _{0}/2}^{t}dt^{\prime
}e^{-iS\left( \mathbf{k}_{0},t^{\prime },t\right) }E\left( t^{\prime
}\right) D_{\mathrm{tr}}\left( \mathbf{k}_{0}+\mathbf{A}\left( t^{\prime
}\right) \right)
\end{equation*}%
\begin{equation}
\times \left[ 1-2\mathcal{N}(\mathbf{k}_{0},t^{\prime })\right] ,  \label{pk}
\end{equation}%
where 
\begin{equation}
S\left( \mathbf{k}_{0},t^{\prime },t\right) =\int_{t^{\prime }}^{t}\left[ 
\mathcal{E}_{eh}\left( \mathbf{k}_{0}+\mathbf{A}\left( t^{\prime \prime
}\right) \right) \right] dt^{\prime \prime }  \label{clac}
\end{equation}%
is the classical action. As is seen, the electron-hole creation amplitude is
defined by the singularity of the transition dipole moment $D_{\mathrm{tr}%
}\left( \mathbf{k}_{0}\right) $ near the Dirac points. In this case due to
the vanishing gap, electron-hole creation is initiated by the non-adiabatic
crossing of the valence band electrons through the Dirac points \cite%
{nondiabatic}. In Figs. 4(a) and 4(b), we plot $\mathcal{N}(\mathbf{k}%
_{0},t) $ for two particular values of crystal momentum $\mathbf{k}_{0}$
along with time-dependent kinetic momentum $\mathbf{k}\left( t\right) =%
\mathbf{k}_{0}+\mathbf{A}\left( t\right) $. The horizontal line is the Dirac
moment $1/\sqrt{3}$ in the units of $k_{b}$. As is clearly seen from Figs.
4(a) and 4(b), the excitation takes place almost instantly when the kinetic
momentum approaches to Dirac point: $\mathbf{k}\left( t\right) =\mathbf{k}%
_{D}$. Thus, we can argue that to excite electrons at any point of BZ it is
necessary, but as we will see is not sufficient to pass through the Dirac
point. The non-adiabatic dynamics suggests to simplify Eq. (\ref{pk})
considerably. As in Ref. \cite{nondiabatic}, we model singularity of the
transition dipole moment defining $D_{\mathrm{tr}}\left( \mathbf{k}%
_{0}\right) $ as a Dirac delta function in both valleys: $D_{\mathrm{tr}%
}\left( \mathbf{k}_{0}\right) =D_{0}\left( \delta \left( \mathbf{k}_{0}-%
\mathbf{k}_{D}\right) -\delta \left( \mathbf{k}_{0}+\mathbf{k}_{D}\right)
\right) $. With this replacement in Eq. (\ref{pk}) we get: 
\begin{equation*}
\mathcal{P}(\mathbf{k}_{0},t)\varpropto i\sum\limits_{t_{d_{+}}}e^{-iS\left( 
\mathbf{k}_{0},t_{d_{+}},t\right) }E\left( t_{d_{+}}\right)
\end{equation*}%
\begin{equation}
-i\sum\limits_{t_{d_{-}}}e^{-iS\left( \mathbf{k}_{0},t_{d_{-}},t\right)
}E\left( t_{d_{-}}\right) ,  \label{pk2}
\end{equation}%
where, for a given $\mathbf{k}_{0}$, $t_{d_{+}}$ and $t_{d_{-}}$ are the
solution of equations:%
\begin{equation}
\mathbf{k}_{0}=\mathbf{k}_{D}-\mathbf{A}\left( t_{d_{+}}\right) ,
\label{plus1}
\end{equation}%
\begin{equation}
\mathbf{k}_{0}=-\mathbf{k}_{D}-\mathbf{A}\left( t_{d_{-}}\right) .
\label{minus1}
\end{equation}

These equations may have several solutions. In Figs. 4(a) and 4(b) we have
two solutions $t_{d_{1+}}$, $t_{d_{2+}}$. Depending on the phase difference $%
\delta S=S\left( \mathbf{k}_{0},t_{d_{1+}},t_{d_{2+}}\right) $ we will have
constructive or destructive interference. Thus, we can argue that to excite
electrons at any point of BZ it is necessary and sufficient to pass through
the Dirac point with constructive interference of multiple passages. This
finding explains interference patterns in Fig. 3. Now let us calculate the
electron populations around respective valleys with the same ansatz. With
the help of Eq. (\ref{pk}), from Eq. (\ref{1}) we have 
\begin{equation*}
N_{+}(\tau _{0}/2)\varpropto \int_{-\tau _{0}/2}^{\tau _{0}/2}dt^{\prime
}E^{2}\left( t^{\prime }\right) -\mathrm{Re}\sum\limits_{t_{2}}\int_{-\tau
_{0}/2}^{\tau _{0}/2}dt_{1}E\left( t_{1}\right)
\end{equation*}%
\begin{equation}
\times E\left( t_{2}\right) e^{iS\left( \mathbf{k}_{D}-\mathbf{A}\left(
t_{1}\right) ,t_{2},t_{1}\right) }  \label{Nplus}
\end{equation}%
where for the given $t_{1}$, $t_{2}$ is defined from equation 
\begin{equation}
\mathbf{A}\left( t_{2}\right) -\mathbf{A}\left( t_{1}\right) =-2\mathbf{k}%
_{D}.  \label{minus}
\end{equation}%
Analogously we have 
\begin{equation*}
N_{-}(\tau _{0}/2)\varpropto \int_{-\tau _{0}/2}^{\tau _{0}/2}dt^{\prime
}E^{2}\left( t^{\prime }\right) -\mathrm{Re}\sum\limits_{t_{2}}\int_{-\tau
_{0}/2}^{\tau _{0}/2}dt_{1}E\left( t_{1}\right)
\end{equation*}%
\begin{equation}
\times E\left( t_{2}\right) e^{iS\left( -\mathbf{k}_{D}-\mathbf{A}\left(
t_{1}\right) ,t_{2},t_{1}\right) }  \label{Nminus}
\end{equation}%
with 
\begin{equation}
\mathbf{A}\left( t_{2}\right) -\mathbf{A}\left( t_{1}\right) =2\mathbf{k}%
_{D},  \label{plus}
\end{equation}

\begin{figure*}[tbp]
\includegraphics[width=1.0\textwidth]{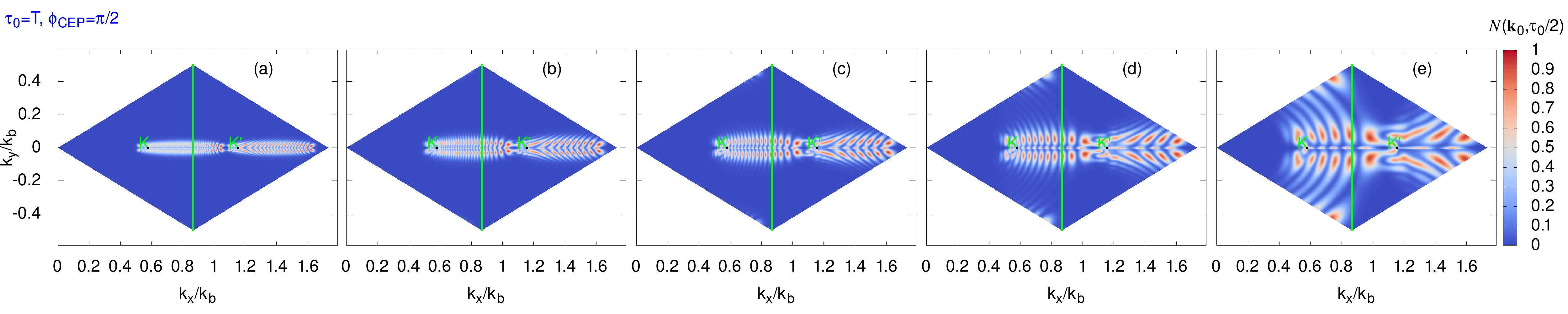}
\caption{Particle distribution function $\mathcal{N}\left( \mathbf{k}%
,t\right) $ (in arbitrary units) at the end of the interaction with a single
cycle sine-pulse ($\protect\phi _{\mathrm{CEP}}=\protect\pi /2$) for
graphene, as a function of scaled dimensionless momentum components ($%
k_{x}/k_{b}$, $k_{y}/k_{b}$). The wave-particle interaction parameter is
taken to be $\protect\chi _{0}=3.5$. (a) for $\protect\omega _{0}=0.1\ 
\mathrm{eV}/\hbar $, (b) for $\protect\omega _{0}=0.2\ \mathrm{eV}/\hbar $,
(c) for $\protect\omega _{0}=0.3\ \mathrm{eV}/\hbar $, (d) for $\protect%
\omega _{0}=0.5\ \mathrm{eV}/\hbar $, and (e) for $\protect\omega _{0}=1.0\ 
\mathrm{eV}/\hbar $.}
\end{figure*}

As is seen from Eqs. (\ref{Nplus}) and (\ref{Nminus}) for both valleys there
is a term which is defined by the intensity area of the pulse. Then we have
a terms which are nonzero when electrons created near the one Dirac point
reach other Dirac point in the momentum space. This explain the threshold
nature of the valley polarizations. The solution of Eqs. (\ref{minus}) and (%
\ref{plus}) are deployed in Fig. 4(c) and 4(d) for the cosine and sine
pulses, respectively. As is seen for cosine pulse if $\left(
t_{1},t_{2}\right) $ pair is the solution, then we have time reversal
solutions $\left( -t_{1},-t_{2}\right) $ and from $\mathcal{E}_{eh}\left( 
\mathbf{k}_{0}\right) =\mathcal{E}_{eh}\left( -\mathbf{k}_{0}\right) $
follows $S\left( \mathbf{k}_{D}-\mathbf{A}\left( t_{1}\right)
,t_{2},t_{1}\right) =-S\left( -\mathbf{k}_{D}-\mathbf{A}\left( -t_{1}\right)
,-t_{2},-t_{1}\right) $ and we get $N_{+}(\tau _{0}/2)=N_{-}(\tau _{0}/2)$.
That is, valley polarization is zero. The same arguments can be made for
infinite pulses as well. For the sine pulse the solutions for the two
valleys Fig. 4(d) are not symmetric and $N_{+}(\tau _{0}/2)\neq N_{-}(\tau
_{0}/2)$.

Having established the excitation dynamics of BZ with the singular
transition dipole moment that works primarily along the $\Gamma -K$\
direction, we next turn to the examination of excitation dynamics in the
perpendicular direction. To this end, we consider the excitation of BZ with
a single cycle sine-pulse ($\phi _{\mathrm{CEP}}=\pi /2$) of the same
interaction parameter $\chi _{0}$\ but at different frequencies. In Fig. 5,
the particle distribution function $N\left( \mathbf{k},\tau _{0}/2\right) $\
at the end of the interaction is shown in reduced BZ for various
frequencies. As indicated in this figure, one can trace the correlation
between the laser frequency and the excitation patterns, which for all cases
have interference fringes. In particular, as the wave frequency decreases,
the fringes become denser and thinner in the $y-$direction. The density of
fringes is defined by the phase difference $\delta S=S\left( \mathbf{k}%
_{0},t_{d_{1+}},t_{d_{2+}}\right) $\ of Eq. (\ref{pk2}) considered above.
For the given $\chi _{0}$\ the phase difference $\delta S$\ is inversely
proportional to $\omega _{0}$, which explains the dependence of the number
of fringes on the frequency. As is also clear, the excitation width in the
perpendicular to laser polarization direction strongly depends on the
frequency of the laser pulse. This can be understood from the x-component of
the transition dipole moment, which for small $k_{y}$\ near the Dirac point
is $D_{\mathrm{tr}}^{(x)}\varpropto 1/\left\vert k_{y}\right\vert $. Thus,
the wave-particle interaction term in Eqs. (\ref{1}) and (\ref{2}) is $%
E\left( t\right) D_{\mathrm{tr}}\varpropto \chi _{0}\omega _{0}/\left\vert
k_{y}\right\vert $. Since electron-hole energy $\mathcal{E}_{eh}\varpropto
\left\vert k_{y}\right\vert $, the excitation width can be estimated from
the condition $\mathcal{E}_{eh}\simeq $\ $E\left( t\right) D_{\mathrm{tr}}$\
which gives $\left\vert k_{y}\right\vert \varpropto \sqrt{\chi _{0}\omega
_{0}}$. As has been shown above, the considered setup of valley polarization
works only when the polarization of the laser is along the $\Gamma -K$\
direction and to have a higher degree of control one must confine the BZ
excitation tighter to the $\Gamma -K$\ direction. Hence, the resulting
scaling $\left\vert k_{y}\right\vert \varpropto \sqrt{\chi _{0}\omega _{0}}$%
\ indicates that low frequencies and moderate intensities are preferable for
valley polarization (when the threshold condition $\chi _{0}\sim 4\pi /3$\
is met).\textrm{\ }
\begin{figure*}[tbp]
\includegraphics[width=1.0\textwidth]{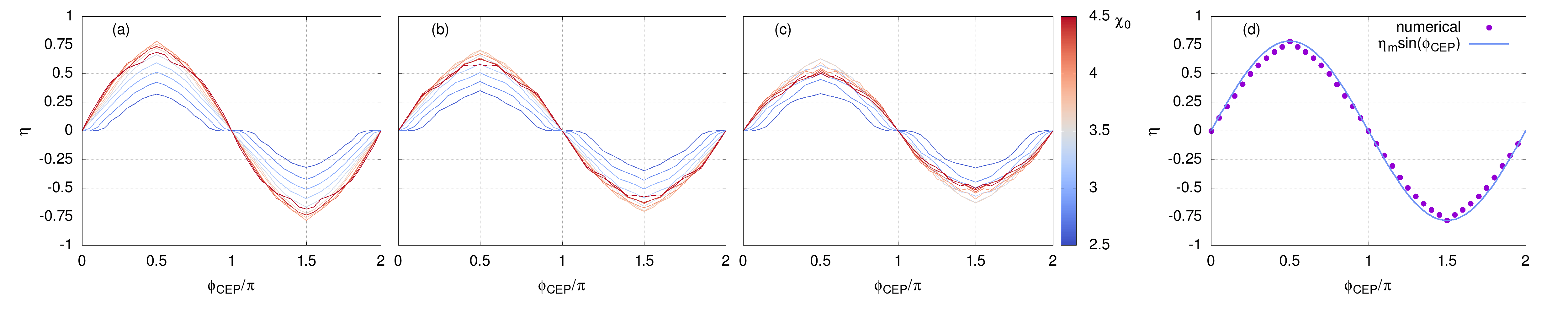}
\caption{Valley polarization for a single-cycle linearly polarized pulse as
a function of carrier-envelope phase for a range of the wave-particle
interaction parameter: (a) for $\protect\omega _{0}=0.1\ \mathrm{eV}/\hbar $%
, (b) for $\protect\omega _{0}=0.2\ \mathrm{eV}/\hbar $, and (c) for $%
\protect\omega _{0}=0.3\ \mathrm{eV}/\hbar $. The color box shows the
strength of the wave-particle interaction parameter. The relaxation time is $%
\Gamma ^{-1}=2\protect\pi /\protect\omega _{0}$. Interpolation of the valley
polarization by the simple harmonic law (c) for $\protect\omega _{0}=0.1\ 
\mathrm{eV}/\hbar $ and $\protect\chi _{0}=4$. }
\end{figure*}
\textrm{\ }
\begin{figure*}[tbp]
\includegraphics[width=1.0\textwidth]{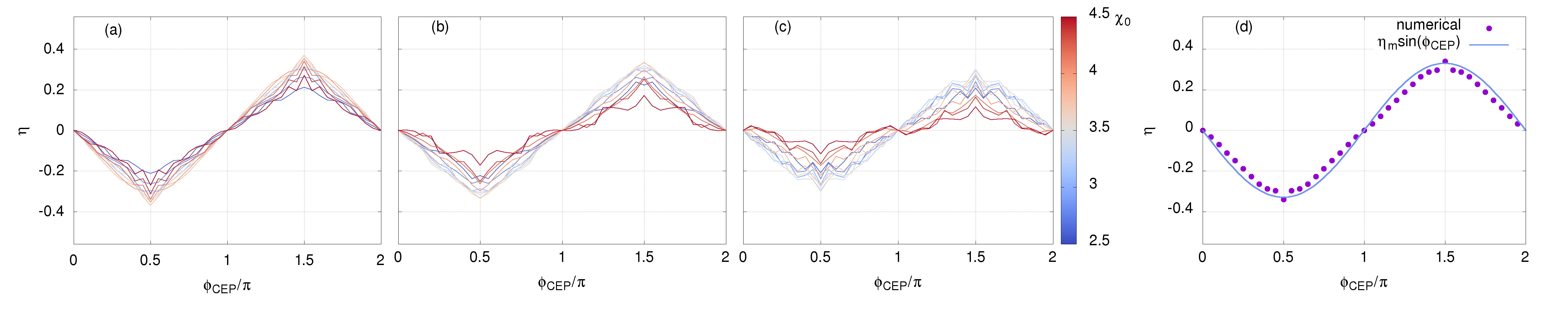}
\caption{ The same as in Fig. (6), but for a two-cycle linearly polarized
pulse. }
\end{figure*}

In Fig. 6 we present the results of our calculations for valley polarization
in a single-cycle linearly polarized pulse as a function of carrier-envelope
phase for a range of the wave-particle interaction parameter and fundamental
frequency. As is seen, in single-cycle pulses we have a strong dependence of
the valley polarization on CEP and we get a rather large valley
polarization. For relatively smaller wave-particle interaction parameter $%
\chi _{0}\simeq 2.5$, the slope $d\eta /d\phi _{\mathrm{CEP}}$\ of the
valley polarization $\eta $\ with respect to the carrier-envelope phase $%
\phi _{\mathrm{CEP}}$\ vanishes when $\phi _{\mathrm{CEP}}/\pi $\ approaches
to an integer. This is the direct manifestation of the threshold nature of
the considered valley polarization scheme: for a relatively smaller
wave-particle interaction parameter the electrons created near the one
valley do not reach the opposite valley for the range of carrier-envelope
phase close to the cosine pulse. Also in Figs. 6(a), 6(b), and 6(c) we see
the result of the negative factor of excitation across the $\Gamma -K$\
direction (cf. Fig. 5). The scaling $\left\vert k_{y}\right\vert \varpropto 
\sqrt{\omega _{0}}$\ indicates that the maximal amplitude of the valley
polarization decreases as the fundamental frequency $\omega _{0}$\
increases. Interestingly, for the wave-particle interaction parameter $\chi
_{0}$\ ranging from $3$\ to $4$, the valley polarization $\eta $\ dependence
on $\phi _{\mathrm{CEP}}$\ tends to harmonic law. Indeed, in Fig. 6(d) we
compare the sinusoidal dependence with the numerically exact result and
obtain fairly good interpolation. Note that these calculations have been
made including relaxation and many-body Coulomb interaction. With the
increase of the pulse duration in Fig. 7, the asymmetric parts in Eqs. (\ref%
{Nplus}) and ((\ref{Nminus}) become smaller and valley polarization
diminished compared with a single-cycle pulse. However, the harmonic law of
valley polarization versus CEP Fig. 7(d) still works. With the increase of
the pulse duration in Figs. 7(a), 7(b), and especially in 7(c), where the
frequency is larger, we see the negative factor of excitation across the $%
\Gamma -K$\ direction because of scaling $\left\vert k_{y}\right\vert
\varpropto \sqrt{\chi _{0}\omega _{0}}$\ which implies that moderate
intensities are preferable for valley polarization. We also see that along
with the harmonic law of the valley polarization, there is a high-frequency
amplitude modulation of the valley polarization, which becomes more
pronounced near its extrema.

\begin{figure*}[tbp]
\includegraphics[width=1.0\textwidth]{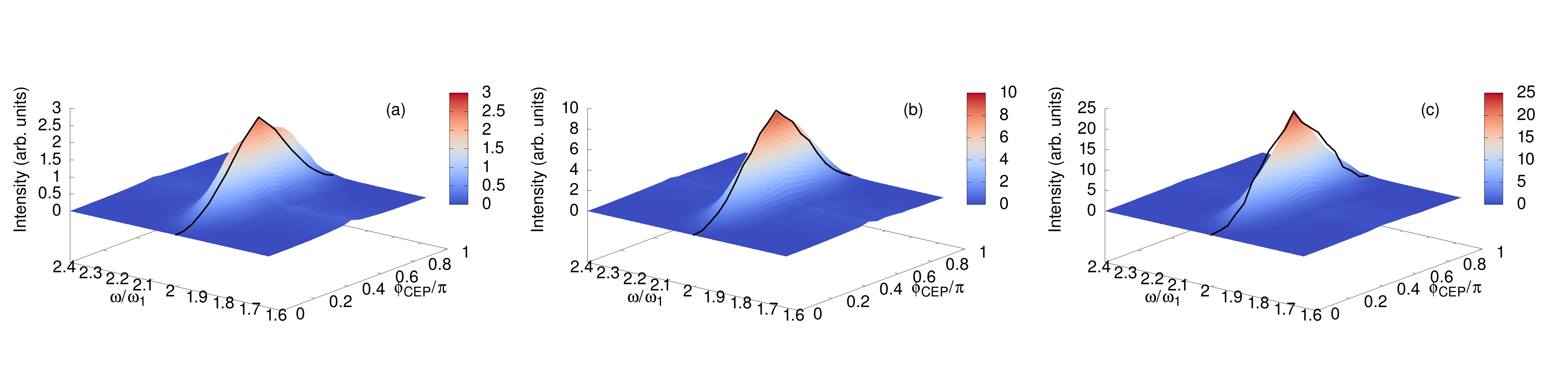}
\caption{The nonlinear response of the graphene in a multi-cycle laser field
of wavelength $800\ \mathrm{nm}$ ($\protect\omega _{1}=1.55\ \mathrm{eV}%
/\hbar $) near the second harmonic via the normalized intensity versus
carrier-envelope phase of a single-cycle linearly polarized preexcitation
pulse: (a) for $\protect\omega _{0}=0.1\ \mathrm{eV}/\hbar $, (b) for $%
\protect\omega _{0}=0.2\ \mathrm{eV}/\hbar $, and (c) for $\protect\omega %
_{0}=0.3\ \mathrm{eV}/\hbar $. The wave-particle interaction parameter for a
preexcitation pulse is taken to be $\protect\chi _{0}=3.5$, while for a
multi-cycle laser field we take $\protect\chi _{1}=0.1\protect\chi _{0}$.
The black (solid) lines over the surface plots are the second harmonic
intensities calculated as $I_{2}\sim \protect\eta ^{2}\left( \protect\phi _{%
\mathrm{CEP}}\right) $.}
\end{figure*}

\subsection{\label{subsec:Semi}Harmonic imaging of valley polarization and
CEP of few-cycle laser pulses in graphene}

Having considered the coherent dynamics of graphene charge carriers
subjected to intense, linearly polarized, few-cycle laser pulses, we then
consider the harmonic mapping of valley polarization and CEP for such
pulses. Although graphene is centrosymmetric, and as a result, there are no
even-order harmonics for equilibrium initial states, it turns out that
spatial dispersion or valley polarization initiates a rather large
second-order nonlinear response \cite{Mer2nd,Golub2,Wehling,Ho}, comparable
to non-centrosymmetric 2D. The sensitivity of the second harmonic to
valley-polarization opens a door for solving two important issues regarding
the valleytronics and light wave electronics: measuring of
valley-polarization and CEP which is important for manipulations with short
laser pulses. Hence, in this subsection we consider harmonic generation in a
multi-cycle laser field of relatively moderate intensity by graphene
preliminary exposed to an intense few-cycle laser pulse.

The EM response in 2D hexagonal nanostructure is determined by the total
current density: 
\begin{equation*}
\mathbf{j}\left( t\right) =-\frac{4}{(2\pi )^{2}}\int_{\widetilde{BZ}}d%
\mathbf{k}_{0}\left\{ \mathbf{v}_{c}\left( \mathbf{k}_{0}+\mathbf{A}\right) 
\mathcal{N}\left( \mathbf{k}_{0},t\right) \right. 
\end{equation*}%
\begin{equation}
+\left. \mathrm{Re}\left[ \mathbf{v}_{\mathrm{tr}}^{\ast }\left( \mathbf{k}%
_{0}+\mathbf{A}\right) \mathcal{P}(\mathbf{k}_{0},t)\right] \right\} ,
\label{je}
\end{equation}%

\begin{figure*}[tbp]
\includegraphics[width=1.0\textwidth]{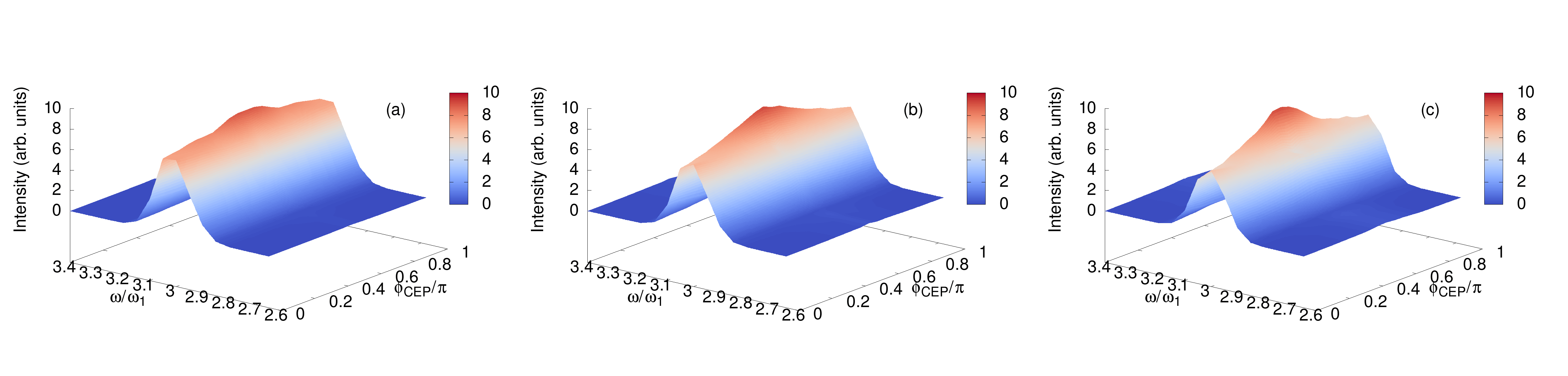}
\caption{The same as in Fig. (8), but for a third harmonic.}
\end{figure*}
\begin{figure}[tbp]
\includegraphics[width=0.4\textwidth]{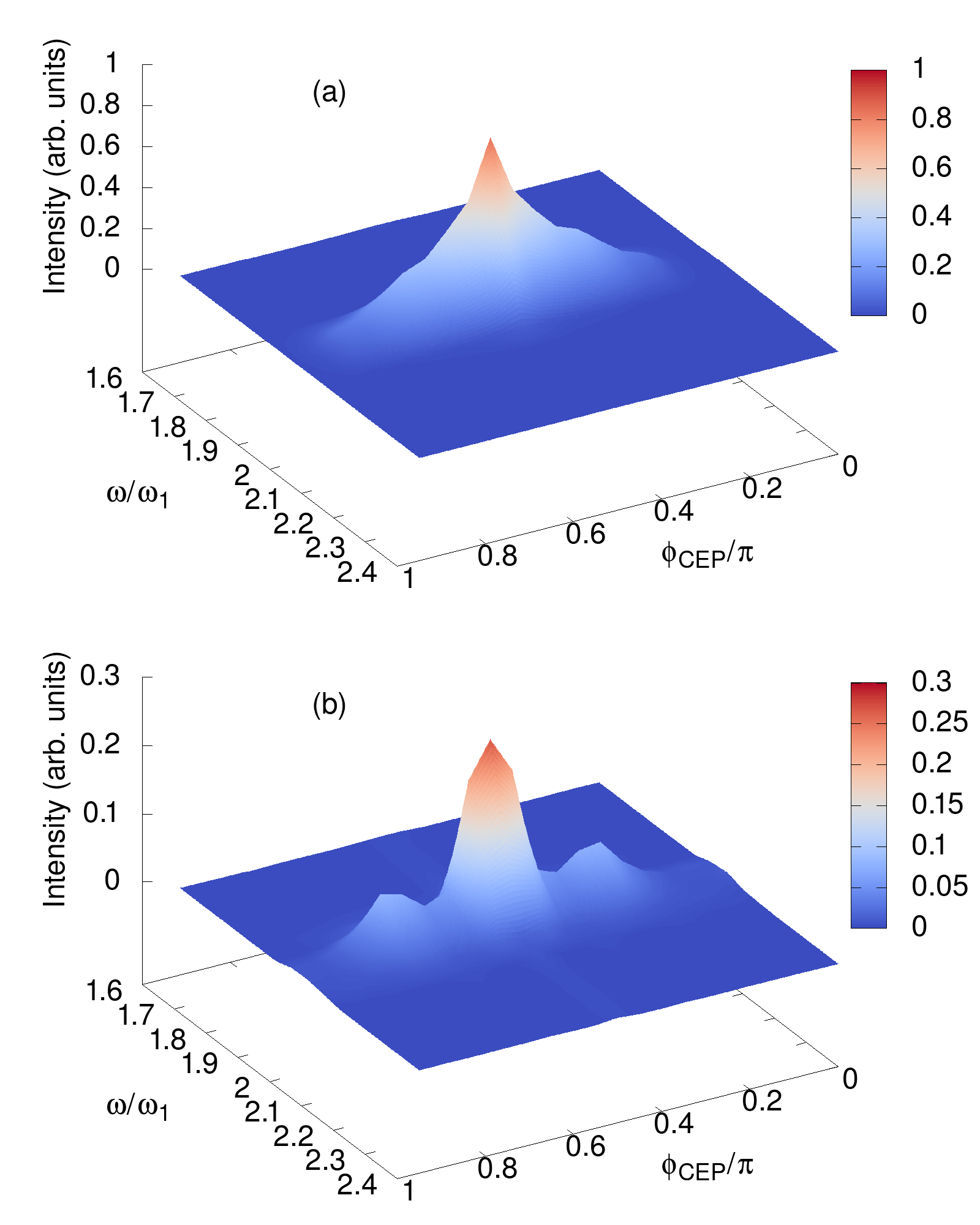}
\caption{The nonlinear response of the graphene in a multi-cycle laser field
of wavelength $800\ \mathrm{nm}$ ($\protect\omega _{2}=1.55\ \mathrm{eV}%
/\hbar $) near the second harmonic via the normalized intensity versus
carrier-envelope phase of a two-cycle (a) and three-cycle (b) linearly
polarized preexcitation pulse. The wave-particle interaction parameters are $%
\protect\chi _{0}=3.5$ and $\protect\chi _{1}=0.1\protect\chi _{0}$, while
the fundamental frequency is $\protect\omega _{0}=0.2\ \mathrm{eV}/\hbar $.}
\end{figure}

where the band velocity $\mathbf{v}_{c}\left( \mathbf{k}\right) =\partial
E\left( \mathbf{k}\right) /\partial \mathbf{k}$ defines intraband
contribution, while transition matrix element $\mathbf{v}_{\mathrm{tr}%
}\left( \mathbf{k}\right) =2iE\left( \mathbf{k}\right) D_{\mathrm{tr}}\left( 
\mathbf{k}\right) \ $of the velocity defines interband contribution. The
Brillouin zone is also shifted to $\widetilde{BZ}=BZ-A$. Note that for the
initially dopped system one should take into account the next-nearest
neighbor hopping term ($\gamma ^{\prime }$) which breaks the electron-hole
symmetry \cite{kretinin2013quantum} and can be crucial for the second
harmonic generation in the presence of asymmetric Fermi energies of valleys 
\cite{wehling2015probing}. Since we consider an undoped system in
equilibrium neglecting thermal occupations, and $\gamma ^{\prime }$\ does
not change the energy difference between the bands, the contribution from $%
\gamma ^{\prime }$\ to the total current density vanishes \cite%
{kretinin2013quantum} when the difference is computed in Eq. (\ref{je}). For
sufficiently large 2D sample, the generated electric field far from the
hexagonal layer is proportional to the surface current: $\mathbf{E}%
^{(g)}(t)=-2\pi \mathbf{j}\left( t\right) /c$ \cite{Mer18}. The HHG spectral
intensity is calculated from the fast Fourier transform of the generated
field $\mathbf{E}^{(g)}(\omega )$. Now we solve Bloch equations Eqs. (\ref{1}%
) and (\ref{2}) with the initial conditions $\mathcal{P}(\mathbf{k},0)=%
\mathcal{P}(\mathbf{k},\tau _{0}/2)$ and $N(\mathbf{k},0)=N(\mathbf{k},\tau
_{0}/2)$, where $\mathcal{P}(\mathbf{k},\tau _{0}/2)$ and $N(\mathbf{k},\tau
_{0}/2)$ are interband polarization and particle distribution function at
the end of the preexcitation pulse.

In Fig. 8, the nonlinear response of the graphene in a multi-cycle laser
field $\tau _{1}=40\pi /\omega _{1}$ of wavelength $800\ \mathrm{nm}$ ($%
\omega _{1}=1.55\ \mathrm{eV}/\hbar $) near the second harmonic via the
normalized intensity versus carrier-envelope phase of a single-cycle
linearly polarized preexcitation pulse is displayed for several fundamental
frequencies. The wave-particle interaction parameter for a multi-cycle laser
field is taken to be smaller by one order compared to the preexcitation
pulse. We specifically consider spectral intensity near the second harmonic
to show that the second harmonic intensity is a robust observable that
provides a gauge of CEP. As is seen from these figures, the second harmonic
signal vanishes for a cosine pulse $\phi _{\mathrm{CEP}}=0$ and for the sine
pulse $\phi _{\mathrm{CEP}}=\pi /2$ reaches maximum values, between these
values it is a monotonic function. In contrast to the second harmonic, the
third harmonic is almost independent of CEP, cf. Fig. 9. That is, for
diagnostic tools the second harmonic is unique. Moreover, as follows from
perturbation theory \cite{Golub2}, when valley polarization is modeled via
different Fermi energies in the $K$ and $K^{\prime }$ valleys, the intensity
of the second harmonic is proportional to the square of valley polarization.
In Fig. 8, we see that the second harmonic intensities calculated as $%
I_{2}\sim \eta ^{2}\left( \phi _{\mathrm{CEP}}\right) \sim \sin ^{2}\left(
\phi _{\mathrm{CEP}}\right) $, where $\eta \left( \phi _{\mathrm{CEP}%
}\right) $ is from Fig. 6 fairly well coincide with the exact numerical
results. Thus, for a single-cycle laser pulse we can reach controllable
valley polarization, which harmonically depends on the carrier-envelope
phase, and vice versa, via valley-polarization we can measure CEP which is
essential for short pulse manipulations and light-wave electronics. This
finding is also valid for two-cycle laser pulse Fig. 10(a). However, for
three-cycle laser pulse Fig. 10(b) the second harmonic intensity is not a
monotonic function in the range $\phi _{\mathrm{CEP}}\subset \left[ 0,\pi /2%
\right] $ and a simple dependence $I_{2}\sim \eta ^{2}$ does not hold. This
is connected with the above-mentioned fact of amplitude modulation of valley
polarization: with the further increase of the pulse duration valley
polarization diminished and becomes comparable to the depth of the amplitude
modulation caused by excitation of BZ across the $\Gamma -K$ direction. Note
that harmonic dependence essentially simplifies the possibility of
measurement of valley polarization and CEP but is not mandatory for their
measurement. For harmonic law one needs one triple of quantities $I_{2},\eta
\rightarrow \phi _{\mathrm{CEP}}$, otherwise one needs several sets of
triples for calibration.

\section{Conclusion}

We have investigated the coherent dynamics of graphene-charged carriers
exposed to an intense few-cycle linearly polarized laser pulse. Solving the
corresponding generalized semiconductor Bloch equations numerically in the
Hartree-Fock approximation taking into account many-body Coulomb
interaction, we demonstrated that valley polarization is strongly dependent
on the CEP, which for a range of intensities can be interpolated by the
simple harmonic law. The obtained numerical results are supported by
approximate and transparent analytical results. Then, we have considered
harmonic generation in a multi-cycle laser field by graphene pre-excited by
an intense few-cycle laser pulse. The obtained results show that valley
polarization and CEP have their unique footprints in the second harmonic
signal, which will allow us to measure both physical quantities for pulse
durations up to two optical cycles, which are vital for valleytronics and
light-wave electronics.

\begin{acknowledgments}
The work was supported by the Science Committee of Republic of
Armenia, project No. 21AG-1C014.
\end{acknowledgments}

\bibliographystyle{apsrev4-1}
\bibliography{bibliography}

\end{document}